\documentclass[useAMS,usenatbib]{mn2e}
\usepackage{graphicx}
\usepackage{epsfig}

\title[{\it Suzaku} observation of CTB 37A (G348.5+0.1)]{{\it Suzaku} observation of Galactic supernova remnant CTB 37A (G348.5+0.1)}
\author[A.~Sezer, F.~G\"{o}k, M.~Hudaverdi and E.N.~Ercan]{A.~Sezer,$^{1,2}$\thanks{E-mail: aytap.sezer@uzay.tubitak.gov.tr
(AS); gok@akdeniz.edu.tr (FG); murat.hudaverdi@uzay.tubitak.gov.tr
(MH); ercan@boun.edu.tr (ENE).} F.~G\"{o}k,$^{3}$
M.~Hudaverdi$^{1}$ and
E.N.~Ercan$^{2}$\footnotemark[1]\thanks{This file has been amended
to highlight the proper use of \LaTeXe\ code with the class file.
These changes are for illustrative purposes and do not reflect the
original paper by A.~Sezer.}\\
$^{1}$T\"UB\.ITAK Space Technologies Research Institute, ODTU
Campus, Ankara, 06531, Turkey\\
$^{2}$Bo\~gazi\d{c}i University, Faculty of Art and Sciences,
Department of Physics, \.Istanbul, 34342, Turkey\\
$^{3}$Akdeniz University, Faculty of Sciences, Department
of Physics, Antalya, 07058, Turkey\\
}
\begin{document}

\date{}

\pagerange{\pageref{firstpage}--\pageref{lastpage}} \pubyear{2011}

\maketitle

\label{firstpage}

\begin{abstract}
We present here the results of the observation of CTB 37A obtained
with the X-ray Imaging Spectrometer onboard the {\it Suzaku}
satellite. The X-ray spectrum of CTB 37A is well fitted by two
components, a single-temperature ionization equilibrium component
(VMEKAL) with solar abundances, an electron temperature of
$kT_{\rm e}\sim0.6$ keV, absorbing column density of $N_{\rm
H}\sim3\times10^{22}$ ${\rm cm^{-2}}$ and a power-law component
with photon index of $\Gamma$ $\sim 1.6$. The X-ray spectrum of
CTB 37A is characterized by clearly detected K-shell emission
lines of Mg, Si, S, and Ar. The plasma with solar abundances
supports the idea that the X-ray emission originates from the
shocked interstellar material. The ambient gas density, and age of
the remnant are estimated to be $\sim$1$f^{-1/2}$${\rm cm^{-3}}$
and $\sim$3$\times10^{4}f^{1/2}$ yr, respectively. The
center-filling X-ray emission surrounded by a shell-like radio
structure and other X-ray properties indicate that this remnant
would be a new member of mixed-morphology supernova remnant class.

\end{abstract}


\section{Introduction}
 CTB 37A (also called G348.5+0.1, $\rmn{RA}(2000)=17^{\rmn{h}} 14^{\rmn{m}} 06^{\rmn{s}}$,
$\rmn{Dec.}~(2000)=-38\degr 32\arcmin$) was discovered by
\citet{b5} in the radio band and has a shell-type morphology with
an angular size of 15 arcmin. From Very Large Array observations
at wavelengths of 6, 20, and 90 cm, \citet{b6} reported that this
supernova remnant (SNR) is expanding in an inhomogeneous region,
and is part of a complex composed of three SNRs: CTB 37A,
G348.7+0.3 (also called CTB 37B), and G348.5-0.0. From the {\it
ASCA} Galactic plane survey data of G348.5+0.1, \citet{b28} found
that the X-ray spectra of the SNR was heavily absorbed by
interstellar matter with $N_{\rm H}\sim2\times10^{22}$ ${\rm
cm^{-2}}$ and the size of X-ray emission was comparable to its
radio structure. \citet{b31} detected OH masers with velocities at
about 20 and 60 km s$^{-1}$, in the direction of CTB 37A at 1720
MHz. \citet{b53} surveyed the environment of this remnant with
associated OH 1720 MHz masers in the CO J=1-0 transition with the
12 Meter Telescope of the NRAO. They reported that a number of
molecular clouds are interacting with the SNR shock fronts.
\citet{b20} using {\it Chandra} and {\it XMM-Newton} data showed
the presence of thermal X-rays from the Northeast part, an
extended non-thermal X-ray source, CXOU J171419.8-383023 in the
Northwest part and a ${\gamma}$-ray source, HESS J1714-385,
coincident with the remnant. They found a high absorbing column
density of $N_{\rm H}\sim3\times10^{22}$ ${\rm cm^{-2}}$. They
claimed that the observed X-ray morphology was a result of
interaction with the inhomogeneous medium surrounding the remnant
and this inhomogeneity was also responsible for the break-out
radio morphology. \citet{b51} using observations with the
Fermi-LAT, have revealed ${\gamma}$-ray emission from this SNR;
the spectrum of the source coincident with CTB 37A was fitted by a
power-law (PL) model with an exponential cutoff at energy $E_{\rm
cut}$=4.2 GeV. Considering the lack of evidence for contribution
by a pulsar and the presence of maser emission for the remnant
they proposed that ${\gamma}$-rays result from SNR-molecular
clouds interactions.

 The distance to the CTB 37A has been estimated from 21 cm
absorbtion measurement to be in between 6.7-13.7 kpc by
\citet{b52}. From velocity measurements of molecular clouds
associated with the remnant, \citet{b53} adopted a distance of
11.3 kpc. So we will use d=11.3 kpc for our calculations
throughout this work.

The location of CTB 37A containing OH maser sources \citep{b31},
1FGL J1714.5-3830 \citep{b51}, X-ray source CXOU J171419.8-383023,
and HESS J1714-385 \citep{b20}, as well as two additional SNRs
(G348.5-0.0 and G348.7+0.3) make this SNR interesting and
important. In addition, its morphology implies a similarity to the
recently proposed group of mixed-morphology (MM) SNRs, increasing
its importance. High quality imaging and spectra obtained from the
data provided by X-ray observatory {\it Suzaku} are used to
produce the results in this work.

The organization of this paper is as follows: we describe the {\it
Suzaku} X-ray observation of CTB 37A, including details of data
reduction in Section 2. The image and spectral analysis are given
in Sections 3 and 4, respectively. Finally, considering its
morphology and investigating radial variation of the electron
temperature, we discuss the implications of MM class, the nature
of thermal and non-thermal components of CTB 37A in Section 5.

\section[]{Observation and Data reduction}

{\it Suzaku} \citep{b17} is the fifth Japanese X-ray astronomy
satellite launched on 2005 July 10. {\it Suzaku} has observed CTB
37A on 2010 February 20. The observation ID and exposure time are
504097010 and 53.8 ks, respectively. The X-ray Imaging
spectrometer (XIS; \citet{b12}) consists of four sets of X-ray CCD
camera system (XIS0, 1, 2 and 3). XIS0, 2, and 3 have
front-illuminated (FI) sensor and provide coverage over the energy
range 0.4$-$12 keV, while XIS1 has a back-illuminated (BI) sensor
providing greater sensitivity at lower energies (0.2$-$12 keV).
The XIS has a field of view (FOV) of 17.8$\times$17.8 arcmin$^{2}$
($1024\times1024$ pixels). Each XIS CCD has an $^{55}$Fe
calibration source, which can be used to calibrate the gain and
test the spectral resolution of data taken using this instrument.
The XIS2 sensor was available only until 2006, therefore, we use
data of XIS0, XIS1, XIS3.

Reduction and analysis of the data were performed following the
standard procedure using the {\sc headas} software package of
version 6.5 and spectral fitting was performed with {\sc xspec}
version 11.3.2 \citep{b2}. The XIS was operated in the normal
full-frame clocking mode, with the standard $3\times3$ and
$5\times5$ editing mode. We generated XIS response matrices using
the {\sc xisrmfgen} software, which takes into account the
time-variation of the energy response. As for generating ancillary
response files (ARFs), we used {\sc xissimarfgen} \citep{b7}. The
latest version of the relevant {\it Suzaku} CALDB files were also
used.

\section{Image Analysis}

Figure 1 shows XIS1 image in 0.3$-$10 keV energy band. We
extracted the spectrum from the brightest region represented with
the outermost solid circle centered at
$\rmn{RA}(2000)=17^{\rmn{h}} 14^{\rmn{m}} 30^{\rmn{s}}$,
$\rmn{Dec.}~(2000)=-38\degr 32\arcmin 07\arcsec$ with a radius of
5.5 arcmin. To derive the radial variation of the electron
temperature $kT_{\rm e}$, we take four apertures with sizes of
0$-$1.5, 1.5$-$2.5, 2.5$-$3.5, 3.5$-$4.5 arcmin. The extended
non-thermal X-ray source (CXOU J171419.8-383023) is excluded from
spectral analysis as shown in Fig. 1. The dashed black circle
centered at ($\rmn{RA}(2000)=17^{\rmn{h}} 14^{\rmn{m}}
19^{\rmn{s}}$, $\rmn{Dec.}~(2000)=-38\degr 24\arcmin 36\arcsec$
with radius 1.7 arcmin) represents the background region. The
lower left corner in the FOV that contains calibration source
emission is also extracted.

Figure 2 shows the XIS0 image in 0.3$-$10 keV energy band, which
is overlaid with the radio image  obtained at 843 MHz by
\citet{b25} for comparison. Dark crosses indicate the direction of
the detected OH (1720 MHz) maser emission at velocities $\sim$
$-65$ km s$^{-1}$ associated with CTB 37A, and the white crosses
show the positions of maser emission at velocities $\sim$ $-22$ km
s$^{-1}$ \citep{b31}. The diamond indicates the position of CXOU
J171419.8-383023 and the circle represents the location of HESS
J1714-385 \citep{b20}.

\section{Spectral Analysis}  XIS spectra were extracted using {\sc
xselect} version 2.4a from all the XISs with a circular extraction
region of radius 5.5 arcmin and are grouped with a minimum of 50
counts bin$^{-1}$. We fit the spectra with a collisional
ionization equilibrium (CIE) model with variable abundances ({\sc
xspec} model ``VMEKAL''; \citet {b15,b16,b13}) modified by
interstellar absorbtion (wabs in {\sc xspec}, \citet {b18}). The
parameters of the absorbing column density ($N_{\rm H}$) and
electron temperature ($kT_{\rm e}$) are set free while all
elements were fixed at solar abundances \citep {b1}. The
best-fitting reduced $\chi^{2}$/d.o.f. for this model is
2222.5/832= 2.67. To find out if there was any contribution from
non-thermal emission we added a PL  component (VMEKAL+PL) yielding
a better reduced $\chi^{2}$ of 935.5/830 = 1.13. Then, Mg, Si, S
and Ar lines in the spectrum were set free, while the rest were
fixed at their solar values, we found insignificant  improvement
in $\chi^{2}$ value (891.6/826 = 1.08). Therefore, we decided to
fix the abundances of Mg, Si, S, and Ar at their solar values. In
Table 1, we present the best-fitting parameters and the statistics
obtained with an absorbed VMEKAL+PL with corresponding errors at
90 per cent confidence level (2.7 $\sigma$). Figure 3 shows the
spectra of the XIS0, XIS1, and XIS3 simultaneously, in the energy
range of 0.3$-$10 keV that is taken from the region shown by the
solid dark circle (the outermost) presented in Fig. 1. On the
other hand, we performed annular spectral analysis for four
regions shown by the circles in Fig. 1 to be able to derive the
radial temperature variations of CTB 37A. The annular regions are
spaced by 1 arcmin from the innermost circle with radius r=1.5
arcmin. VMEKAL+PL spectral model were also fitted to each annulus,
while fitting them we kept the absorbing column density $N_{\rm
H}$ at its best-fitting value for the entire remnant. Figure 4
shows the electron temperature variations with respect to the
radius.

\begin{table*}
\centering
 \begin{minipage}{140mm}
  \caption{Best-fitting parameters and $\chi^{2}$ values of the spectral fitting in the full energy band (0.3$-$10 keV)
   with an absorbed VMEKAL+PL model with corresponding errors at 90
   per cent confidence level (2.7 $\sigma$).}
 \begin{tabular}{@{}cccc@{}}
  \hline
      Component&Parameters & VMEKAL+PL&\\
 \hline
 Wabs&$N_{\rm H}$($\times10^{22}$$\rm cm^{-2})$ & 2.9 $\pm 0.1$&  \\
VMEKAL &$kT_{\rm e}$(keV) & 0.63 $\pm 0.02$&\\
 Abundance\footnote{(1) indicates that the elemental abundance is fixed at solar
\citep{b1}.}&Mg  & (1)&\\
 &Si & (1)&\\
 &S & (1)&\\
 &Ar & (1)&\\
&VEM\footnote{Volume emission measure VEM=$\int n_{\rm e}n_{\rm
H}$dV in the unit of $10^{58}$ $\rm cm^{-3}$, where $n_{\rm e}$
and $n_{\rm H}$ are number densities of electrons and protons,
respectively, and V is
the X-ray emitting volume.}&72.3 $\pm 2.1$& \\
&Flux\footnote{Unabsorbed flux in the $0.3-10$ keV energy band in the unit of $10^{-9}$ erg $\rm s^{-1}$$\rm cm^{-2}$.}&1.36 $\pm 0.03$ & \\
 PL &Photon Index &1.6 $\pm 0.1$&\\
 &norm($\times10^{-2}$photons $\rm cm^{-2}s^{-1}$) & 1.9 $\pm 0.2$ &\\
 &Flux\footnote{Total unabsorbed flux of the sum of the VMEKAL and PL components in the $0.3-10$ keV energy band in the unit of $10^{-9}$ erg $\rm s^{-1}$$\rm cm^{-2}$.} &1.5 $\pm 0.1$ &\\
 &$\chi^{2}$/d.o.f.  &935.6/830=1.13 & \\
\hline
\end{tabular}
\end{minipage}
\end{table*}

\section{Discussion and Conclusions}

In this work, we provide a description of the X-ray emission of
CTB 37A based on {\it Suzaku} archival data. We obtained a clear
image and high quality spectra of diffuse X-ray emission. We have
examined the thermal and non-thermal emissions coming from the
remnant. The X-ray spectrum of CTB 37A is characterized by thermal
emission dominated by K-shell emission lines of Mg, Si, S, and Ar
which are clearly detected.

As seen from Fig. 2, the radio emission of CTB 37A comprises a
partial shell towards the north and east and an extended outbreak
to the south, while the X-ray emission has a deformation along the
southwest limb, where the morphology appears indented. The reason
for such a deformation may be due to the inhomogeneous medium
along this specific region and this may well be supported by the
fact that remnant is close to the Galactic plane and several OH
masers at 1720 MHz are detected towards CTB 37A \citep {b31}
(shown by crosses). It may also be relevant that $\gamma$-ray
emission thought to be associated with the interaction between CTB
37A and dense surrounding material that has been detected with the
{\it Fermi LAT} \citep {b51}.

\subsection{Implications for mixed-morphology}

SNRs were originally divided into shell-like, Crab-like
(plerionic), and composite (shell-like containing plerions)
remnants \citep {b42} according to their X-ray morphology.
Recently, an additional MM class (also called thermal composite)
appeared, which are center-filled in X-rays and shell-like at
radio wavelengths \citep {b23, b24, b34}. As seen from Fig. 2, CTB
37A has a shell-like morphology in the radio band while
centrally-filled in X-ray band, in this regard, our first
impression is CTB 37A seems to be a MM SNR. Examples of well-known
MM SNRs include W28 \citep {b48}, G290.1-0.8 \citep {b49}, and IC
443 \citep {b4}. \citet {b3} reported important results about this
class by compiling a list of 26 MM SNRs.

The X-ray characteristics of MM remnants have been defined by
\citet {b34} as follows: (1) The radial temperature distribution
is relatively flat; (2) The X-ray emission arises primarily from
shocked interstellar material, and not from ejecta; (3) The
remnants are typically located close to molecular clouds or very
dense regions; and (4) The dominant X-ray emission is thermal in
nature. Subsequent studies indicated that MM SNRs had a complex
plasma structure with multiple components (e.g. \citet {b48}) and
enhanced abundances (e.g. \citet {b50,b49}), and they have evolved
over $\sim$$10^{4}$ yr, which means that the plasma is in CIE or
an overionization condition \citep {b47}. As can be seen from Fig.
4, the annular analysis of CTB 37A may well be indicating a very
small scale radial variation in its temperature between the
selected regions. The best-fitting metal abundances are found to
be solar in general, confirming  the absence of ejecta
contamination in selected regions. It may support the idea that
the X-ray emission originates from the shocked interstellar
material. The plasma of CTB 37A is in a collisional ionization
equilibrium condition and is located in a region with density
variation, possibly associated with molecular clouds. The plasma
has thermal and non-thermal emission, but the emission is
dominated by thermal component ($\sim90$ per cent of the total
X-ray flux). These X-ray properties of CTB 37A exemplify the
typical characteristic of MM SNRs as defined by \citep {b34}.

There are a few models that can produce centrally enhanced X-ray
emission such as  evaporation of clouds left relatively unspoiled
after the passage of the SNR blast wave (e.g. \citet {b9}) and
``fossil'' thermal radiation that is detectable as thermal X-rays
from the hotter interior as the shell of an expanding  SNR cools
below $\sim$$10^{6}$ K and  becomes invisible due to interstellar
absorbtion (e.g. \citet {b42}), as the SNR evolves, the
temperature and density of the hot interior plasma gradually
become uniform through thermal conduction (e.g. \citet {b10}) or
evolution in a medium with a density gradient viewed along the
line of sight \citep {b26}. The evaporation model requires dense
clouds, the thermal conduction model requires a relatively high
density ambient medium. CTB 37A is located in a region of greatly
varying density, with OH maser sources indicating interaction with
molecular clouds. In this regard, the center-filled X-ray
morphology of CTB 37A is consistent with evaporating clouds model.
The radial temperature variation in the plasma of CTB 37A
($kT_{\rm e}$ $\sim$ 0.6-0.8 keV) is consistent with other MM SNRs
such as W44 \citep {b29,b21}, 3C391 \citep {b32,b33} and HB21
\citep {b27}. Very small temperature variation in the plasma of
CTB 37A as shown Fig. 4 can be explained by both evaporation and
thermal conduction models. Future deep X-ray observations and
detailed spectral analysis of this remnant would give more
detailed information to compare with theoretical models that
produce MM SNRs.

\subsection{Thermal component}

The X-ray emission of CTB 37A is dominated by thermal emission
that can be best described by an absorbed CIE plasma model
(VMEKAL) with an absorbing column density of $N_{\rm H}\sim
3\times10^{22}$ $\rm cm^{-2}$, an electron temperature of $kT_{\rm
e}\sim0.6$ keV, and solar abundances of Mg, Si, S, and Ar, which
indicate a shocked interstellar/circumstellar material origin.

For full ionization equilibrium, the ionization timescale,
$\tau=n_{\rm e}t$, is required to be $\geq$ $10^{12}$ $\rm
cm^{-3}$s, where t is the plasma age or the time since the gas was
shock-heated \citep {b14}. To determine the age of the remnant,
$n_{\rm e}$ should be estimated from the emission measure, $n_{\rm
e}n_{\rm H}V$, which is related to the normalization of the VMEKAL
model according to the equation, norm=$n_{\rm e}n_{\rm H}
V$/($4\pi d^{2}$$10^{14}$), where V is the X-ray emitting volume,
$n_{\rm H}$ is the volume density of hydrogen and d is the
distance. For simplicity, we assumed the emitting region to be a
sphere of radius 5.5 arcmin. Considering the possibility that less
than the entire volume is filled, we write the volume V=$V_{\rm
s}f$, where $V_{\rm s}$ is the full spherical volume, $f$ is the
filling factor. We then carry the $f$ factor through our
calculations to show the explicit dependence of each derived
quantity on this factor. Knowing that the SNR is at a distance of
11.3 kpc and $n_{\rm e}=1.2n_{\rm H}$, we estimated the emission
volume to be $V\sim 6.7\times10^{59}f$ ${\rm cm^{3}}$.
Consequently, we find an ambient gas density of $\sim$1$f^{-1/2}$
${\rm cm}^{-3}$ and age of $\sim$$3\times10^{4}f^{1/2}$ yr
(assuming $n_{\rm e}t\sim 1\times10^{12}$ $\rm cm^{-3}$s),
implying that CTB 37A is a middle-aged SNR. Finally, we calculated
total mass of the X-ray emitting plasma, $M_{\rm x}$, by $M_{\rm
x}$=$m_{\rm H}n_{\rm e}V$$\sim 530 f^{1/2}{M\sun}$, where $m_{\rm
H}$ is mass of a hydrogen atom, $\mu$=0.604 is the mean atomic
weight.

\subsection{Power-law component}

The Suzaku X-ray spectral data of CTB 37A is well fitted with a
thermal component and an additional hard component. There could be
a few reasons for the hard X-ray emision: (i) an association with
a classical young pulsar, (ii) a contribution from an extended
non-thermal X-ray source (CXOU J171419.8-383023), (iii)
overionization of the plasma which produces excess hard emission
as has been the case for IC443 \citep {b60}. The hard component is
well fitted by a PL model with a photon index value of $\sim$1.6.
This value is consistent with that of classical young pulsar value
ranging in between 1.1 and 1.7 \citep {b37}. However, there is no
pulsar reported that is associated with this remnant. In the
Northwest region of the CTB 37A, an extended non-thermal X-ray
source (CXOU J171419.8-383023) is reported
($\rmn{RA}(2000)=17^{\rmn{h}} 14^{\rmn{m}} 20^{\rmn{s}}$,
$\rmn{Dec.}~(2000)=-38\degr 30\arcmin 20\arcsec$) by \citet {b20}.
In their work, a non-thermal emission from the source  with a
spectral index of $\sim$1.32  is found which is lower (harder)
than our best-fitting value of $\sim$1.6. Although we excluded
CXOU J171419.8-383023 (with radius $2.1$ arcmin) from our spectra
during our spectral analysis, our fits required a non-thermal
component. To investigate this, we performed spectral analysis
also for individual regions by selecting small rectangular regions
that are being further away from the known extended non-thermal
source. The non-thermal flux is found to be stronger for the
selected small regions nearby the source compared to the ones
further away. We have obtained an unabsorbed flux value of $F_{\rm
x}\sim1.4\times10^{-9}$ erg $\rm s^{-1}$$\rm cm^{-2}$ for the
non-thermal extended source in  the 0.3$-$10 keV energy range.
When we compare it with the unabsorbed flux value ($F_{\rm
x}\sim0.14\times10^{-9}$ erg $\rm s^{-1}$$\rm cm^{-2}$) of PL
component of best-fitting, we find a factor of ten difference
between them. This difference indicates that the extended source
is most likely the origin of the PL component and emission
scattered from the source into the field of the rest of the
remnant by a broad point spread function of the {\it Suzaku}
mirrors.

The spectral studies of CTB 37A indicate that the plasma is best
described by a thermal component in CIE condition with solar
elemental abundances and a non-thermal component with a photon
index of $\sim$1.6. Thermal emission possibly originates  from the
shocked interstellar material with ambient gas density of
$\sim$1$f^{-1/2}$${\rm cm^{-3}}$. The best spectral fits  require
an ionization timescale  of  $\tau\geq$ $10^{12}$ $\rm cm^{-3}$s,
implying an age of $\sim$3$\times10^{4}f^{1/2}$ yr. The origin of
the power-law component is more likely the effect of the
contribution from the extended source (CXOU J171419.8-383023)
located  in the Northwest part of the remnant. CTB 37A is most
likely a new member of mixed-morphology SNR.

\section*{Acknowledgments}
We thank Dr. Patrick Slane for his valuable comments and
suggestions which helped to improve the overall quality of the
manuscript. AS is supported by T\"{U}B\.{I}TAK PostDoctoral
Fellowship. This work is supported by the Akdeniz University
Scientific Research Project Management and by T\"{U}B\.{I}TAK
under project codes 108T226 and 109T092. The authors also
acknowledge the support by Bo\u{g}azi\c{c}i University Research
Foundation under 2010-Scientific Research Project Support (BAP)
project no:5052.

\onecolumn

\begin{figure}
\centering
  \vspace*{17pt}
  \includegraphics[width=8cm]{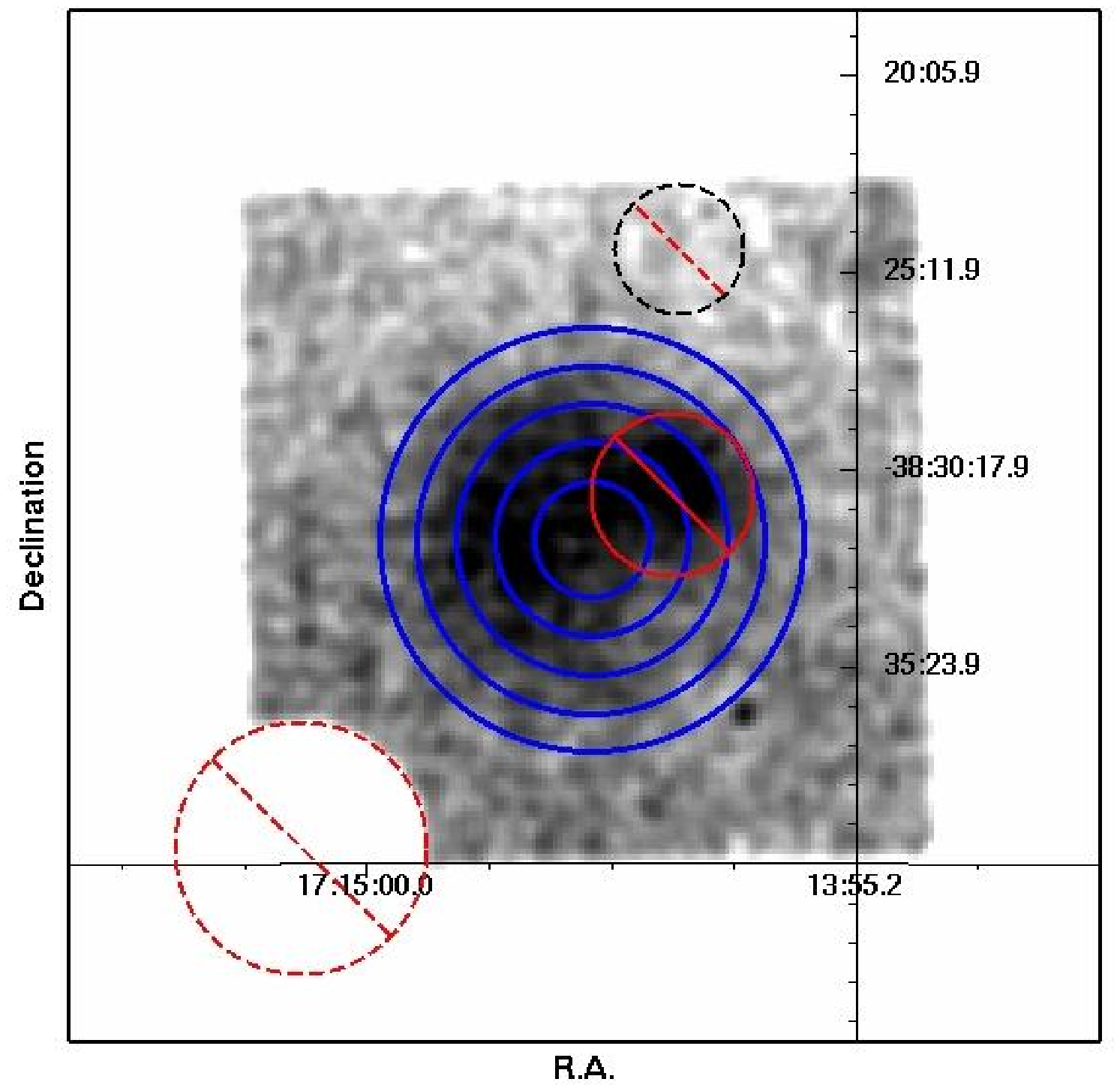}
  \caption{$Suzaku$ XIS1 image of CTB 37A in the 0.3$-$10 keV energy band.
   The spectral integration regions for the source (annular apertures) and the background are indicated
    with the solid dark and dashed black circles, respectively. The X-ray source (CXOU
J171419.8-383023) is extracted from the spectral analysis as shown
in the figure. The corner of the CCD chip illuminated
     by $^{55}$Fe calibration source is excluded from the image. The coordinates (RA and Dec) are
referred to epoch J2000.}
\end{figure}

\begin{figure}
\centering
\vspace*{17pt}
\includegraphics[width=8.5cm]{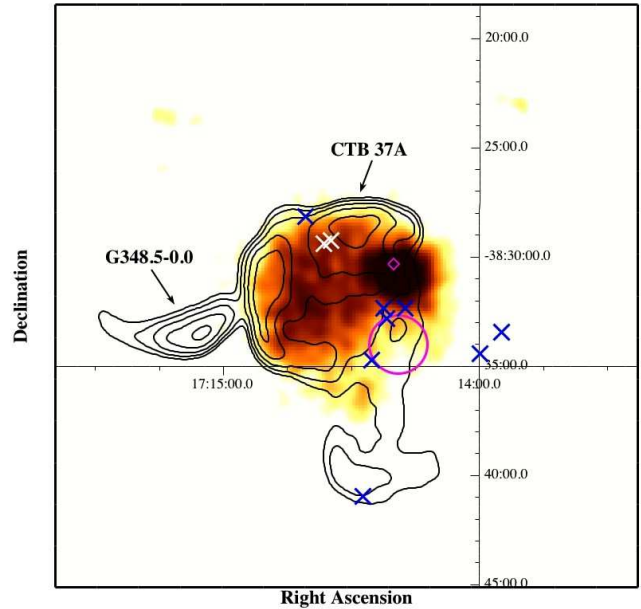}
\caption{CTB 37A is displayed in 0.3$-$10 keV energy band X-ray
data obtained from {\it Suzaku} XIS0 detector. The image is
overlaid by 843 MHz radio isocontours logarithmically scaled
(0.23, 0.36, 0.57, 0.90, 1.42, 2.23 and 3.51 mJy/Beam). Radio
emission clearly shows 2 radio SNRs, CTB 37A and G348.5-0.0. X-ray
data is smoothed with 3 pixel Gaussian in order to highlight the
structure. The colour-coding shows the brightness levels from 1.3
counts/pix to 13.36 counts/pix in logarithmic scale from bright to
dark. The masers from the field are indicated with dark and white
crosses at velocities $\sim$ $-65$ km s$^{-1}$ and $\sim$ $-22$ km
s$^{-1}$, respectively. The diamond shows the extended non-thermal
X-ray source CXOUJ 171419.8-383023. The location of HESS J1714-385
is represented with the circle, in which the positional error is
specified with the radius of 1.2 arcmin. The coordinates (RA and
Dec) are referred to epoch J2000.}
\end{figure}

\begin{figure}
\centering
  \vspace*{17pt}
 \includegraphics[width=15cm]{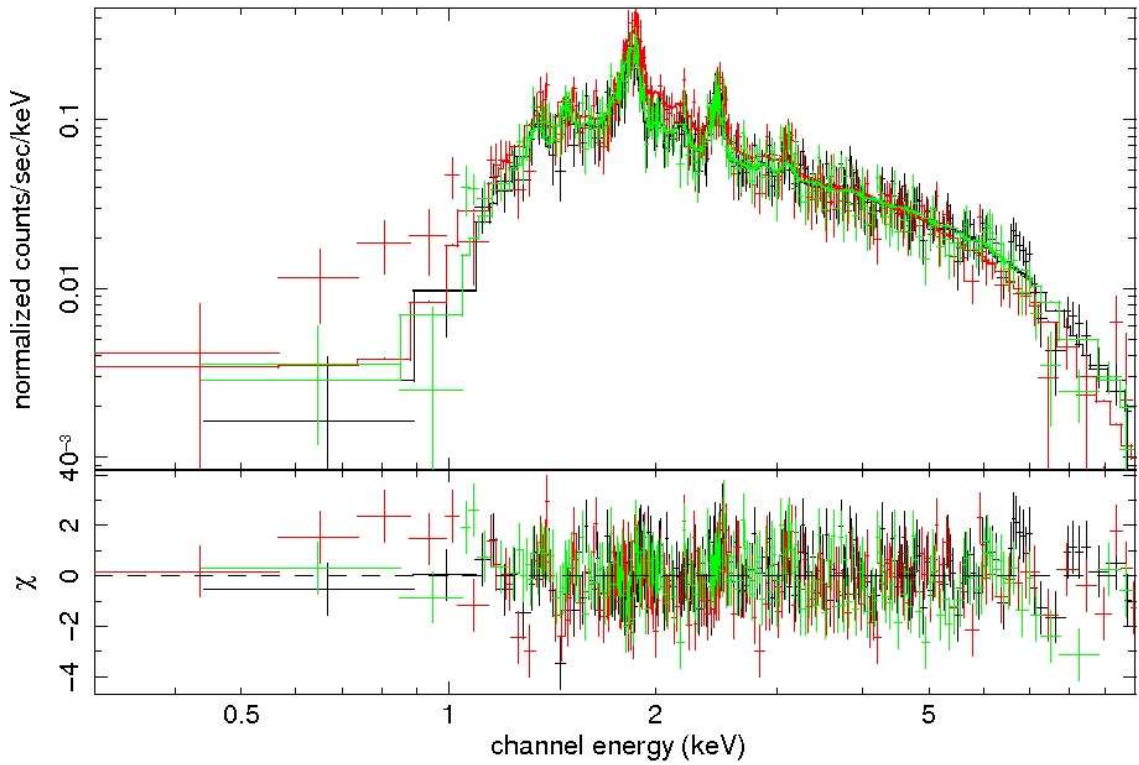}
  \caption{Background-subtracted XIS (XIS1, XIS0 and XIS3) spectra of
 CTB 37A in 0.3$-$10 keV energy band fitted with an
absorbed VMEKAL and PL model. The bottom panel is the residuals
from the best-fitting model.}
\end{figure}

\begin{figure}
\centering
  \vspace*{17pt}
  \includegraphics[width=8.5cm]{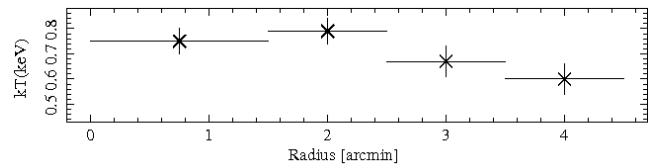}
  \caption{The radial variation of the electron temperature of CTB 37A.}
\end{figure}

\end{document}